\documentclass[12pt]{article}

\catcode`\@=11

\global\arraycolsep=2pt
\usepackage{amsbsy,amssymb,latexsym,amsfonts,amsmath}
\usepackage{graphicx,color}

\begin{document}
\begin{flushright}
\parbox{4.2cm}
{RUP-16-17}
\end{flushright}

\vspace*{0.7cm}

\begin{center}
{ \Large Topologically twisted renormalization group flow  and its holographic dual }
\vspace*{1.5cm}\\
{Yu Nakayama}
\end{center}
\vspace*{1.0cm}
\begin{center}

Department of Physics, Rikkyo University, Toshima, Tokyo 171-8501, Japan

\vspace{3.8cm}
\end{center}

\begin{abstract}
 Euclidean field theories admit more general deformations than usually discussed in quantum field theories because of mixing between rotational symmetry and internal symmetry (a.k.a topological twist). Such deformations may be relevant, and if the subsequent renormalization group flow leads to a non-trivial fixed point, it generically gives rise to a scale invariant Euclidean field theory without conformal invariance. Motivated by an ansatz studied in cosmological models some time ago, we develop a holographic dual description of such renormalization group flows in the context of AdS/CFT. We argue that the non-trivial fixed points require fine-tuning of the bulk theory in general, but remarkably we find that the $O(3)$ Yang-Mills theory coupled with the four-dimensional Einstein gravity in the minimal manner supports such a background with the Euclidean AdS metric.

\end{abstract}

\thispagestyle{empty} 

\setcounter{page}{0}

\newpage

\section{Introduction}
In introductory courses on quantum field theory, the Lorentz invariance is a holy grail and we are not allowed to question its origin. Equally sacred is the unitarity and we never give it up. In the recent analysis of the renormalizationg group flows, the both have played significant roles to uncover their veils,\footnote{See e.g. \cite{Komargodski:2011vj}\cite{Luty:2012ww}\cite{Nakayama:2013is} for the recent developments in four-dimensions.} so it is totally legitimate that we worship them.

However, not all statistical models and Euclidean field theories, which are obtained as their continuum limit, have such features. Indeed, as we will discuss in this paper, Euclidean field theories admit more general deformations than usually discussed in quantum field theories with the  Lorentzian signature. This is because what we mean by the Euclidean rotation is not unique, and it may mix with the internal symmetry. Unlike in quantum field theories with the Lorentzian signature, we are now entitled to and sometimes we are even forced to think about the origin of the Euclidean rotation and unitarity (or reflection positivity).

Let us, for example, take a lattice system as a constructive realization of a Euclidean field theory in the continuum limit. Then it is not so obvious what we mean by the Euclidean rotation because it is broken by the lattice regularization first of all. In some cases, there are various different realizations of the rotational symmetry. The similar situations arise in the condensed matter physics, in particular, in the study of the frustrated spin systems both in the classical case \cite{Kawamura} as well as in the quantum case \cite{qspin}. The rotational symmetry that we would like to expect near the criticality may not be the same as the geometrical rotation of the microscopic Hamiltonian.

In the string theory literature, different possibilities to realize the Euclidean rotation  are often called ``topological twist" \cite{Witten:1988ze}\cite{Witten:1988xj}. Even if we start with a theory with the same symmetry, the topological twist changes its interpretation such as what we mean by the spin quantum number under the Euclidean rotation.\footnote{We understand that in order to obtain topological field theories, we need to define the topological BRST charge in addition to the twisting procedure. In this paper, we use  the word ``topological twist" irrespective of the existence of the BRST charge. } In this case, it is more than the mere interpretation: the idea of topological twist has significant applications. By changing the spin quantum number of operators, one may be able to preserve the symmetry on non-trivial manifold \cite{Witten:1988ze}\cite{Witten:1988xj} or on the lattice, which has been difficult to realize otherwise. One of the most important applications in this directions is how to realize supersymmetric field theories on non-trivial manifold \cite{Witten:1988ze}\cite{Witten:1994ev}\cite{Dumitrescu:2012ha} or on the lattice \cite{Elitzur:1982vh}\cite{Sakai:1983dg}\cite{Kawamoto:1999zn}\cite{Kato:2003ss}\cite{D'Adda:2004jb} (see e.g. \cite{Catterall:2009it} for a review). By using the technique of topological twist, we may be able to make the supercharge singlet under the Euclidean rotation to make it easier to preserve.

More generically, after the topological twist, non-trivial tensors in original Euclidean field theories may become singlet, and we may add such operators to the Lagrangian to deform the theory. Such deformations, however, are more or less neglected in usual quantum field theory  literature because they typically break the unitarity. Nevertheless, in Euclidean field theories, they are as important as non-topologically-twisted scalar deformations that we usually study in the renormalization group analysis.

The topologically twisted scalar deformations may be relevant under the renormalization group flow. Then they may destabilize the criticality and may require further fine-tuning on the lattice. The goal of this paper is to study possible consequences.
We will approach the problem both from the field theory viewpoint and the holographic viewpoint. Our holographic analysis is motivated by cosmological models studied in the literature. We will see that the non-trivial fixed points in general require fine-tuning of the bulk theory. Remarkably, however, we find that the $O(3)$ Yang-Mills theory coupled with the four-dimensional Einstein gravity in the minimal manner supports such a  background. The background provides us with a sought-after example of scale invariant but not-conformal invariant geometry.
The construction seems ubiquitous and we expect novel types of holographic backgrounds to be explored in many supergravity models with the $O(3)$ gauge field.

\section{Topological twist and topologically twisted deformation}
Let us consider a $d$-dimensional Euclidean field theory with $O(d)_{\mathrm{internal}}$ global symmetry, which may be a part of a larger symmetry. By assumption, the original theory has a symmetric conserved energy-momentum tensor as well as the $O(d)_{\mathrm{internal}}$ global current
\begin{align}
\partial^\mu T_{\mu\nu} &= 0 \cr
\partial^\mu J_{\mu}^{[IJ]} & = 0 \ ,
\end{align}
where the internal indices $I, J = 1, \cdots d$ are anti-symmetrized in $[IJ]$ to represent the adjoint of $O(d)_{\mathrm{internal}}$,  and $\mu,\nu = 1,\cdots d$ indices represent the original Euclidean rotation.

The central idea of the topological twist is to define the new rotational symmetry as the diagonal $O(d)_{\mathrm{twisted}}$ of $O(d)_{\mathrm{rotation}} \times O(d)_{\mathrm{internal}}$. Under the new rotational symmetry, various tensor transforms differently than its original representation. We may have new scalar operators under the twisted rotational symmetry   (i.e. singlet under $O(d)_{\mathrm{twisted}}$) and they are what we would like to use to deform the original theory.

After the topological twist, the energy-momentum tensor becomes
\begin{align}
\tilde{T}_{\mu\nu} = T_{\mu\nu} + \frac{1}{2} \left( \partial_\rho J_\mu^{[IJ]} \delta_{I\nu} \delta_J^\rho + \partial_\rho J_{\nu}^{[IJ]} \delta_{I \mu} \delta_{J}^\rho \right)  \ . \label{newEM}
\end{align}
The new energy-momentum tensor is symmetric and conserved, but the additional term shows that the new rotational charge has a contribution from the original global $O(d)_{\mathrm{internal}}$ symmetry in addition to the original Euclidean rotation.
We should note that so far we have done nothing to the original theory under consideration. It still possesses two independent $O(d)$ symmetries. From the continuum Euclidean field theory viewpoint, what we have done is just renaming of various tensors as irreducible representations of $O(d)_{\mathrm{twisted}}$.

Things become non-trivial if we consider the deformation of the theory after the topological twist. Let us introduce a deformation by a topologically twisted scalar operator
\begin{align}
\delta S = \int d^d x O_{\mu\nu\cdots}^{IJ \cdots} \delta_I^{\mu} \delta_J^{\nu}  \cdots \label{td}
\end{align}
The deformation preserves only the diagonal of $O(d)_{\mathrm{twisted}}$ but breaks $O(d)_{\mathrm{rotation}}$ and $O(d)_{\mathrm{internal}}$ separately, so it gives rise to a novel Euclidean system that is only available after the topological twist.
In many statistical systems based on lattice, the rotational symmetry is an emergent symmetry, however, so a priori it is not obvious which combinations of the rotational symmetry and the internal symmetry are compatible with the lattice symmetry. Then we would naturally expect that the deformations such as \eqref{td} are induced on the lattice.

If the original untwisted theory has a reflection positivity, the allowed deformations in \eqref{td} are limited due to the unitarity constraint \cite{Mack:1975je}. In $d>3$, when the spin of the original tensor $O_{\mu\nu \cdots}^{IJ \cdots}$ are higher than two, then such deformations have scaling dimensions greater than $d$ and they are irrelevant under the perturbative renormalization group flow near the undeformed theory. Thus, interesting topologically twisted deformations come from the spin one operators in the fundamental representation of $O(d)_{\mathrm{internal}}$, the spin two operators in the symmetric traceless tensor representation of $O(d)_{\mathrm{internal}}$ or the adjoint operators both in $O(d)_{\mathrm{rotation}}$ and $O(d)_{\mathrm{internal}}$.
 In particular, unless we are working at the free Gaussian fixed point, the latter choice may not be available. 

We note that $d=3$ is special because in $O(3)$, the antisymmetric tensor is equivalent to a vector. Thus the $O(3)_{\mathrm{internal}}$ current operator itself becomes singlet after the topological twist. Since the conserved current in $d=3$ dimensions has the scaling dimension $2$, it always gives rise to a relevant topologically twisted deformation of the UV fixed point with  $O(3)_{\mathrm{internal}}$  symmetry:
\begin{align}
\delta S = \int d^3x \epsilon_{\mu IJ} J^{\mu IJ} \ , 
\end{align}
where $\epsilon_{\mu IJ} = \epsilon_{\mu \nu \rho} \delta^{\nu}_I \delta^{\rho}_J$ is the Levi-Civita tensor.

The topologically twisted deformations so far considered may be relevant under the renormalization group flow. We now argue that if it reaches a non-trivial renormalization group fixed point, it will generically give rise to a scale invariant but not conformal field theory. Suppose the theory  under consideration is conformal invariant before the topologically twisted deformation so that the trace of the original energy-momentum tensor vanishes (or is improved to be zero):
\begin{align}
T^{\mu}_{\mu} = 0 
\end{align}
After the topologically twisted deformation, the trace of the energy-momentum tensor becomes
\begin{align}
\tilde{T}^{\mu}_{\mu} = \beta O_{\mu\nu\cdots}^{IJ \cdots} \delta_I^{\mu} \delta_J^{\nu}  + \partial_\mu J_\rho^{[IJ]} \delta_{I}^\rho \delta_J^\mu \ , 
\end{align}
where $\beta$ is the beta function for the topologically twisted deformation. If the beta function possesses a non-trivial zero in the weakly coupled regime, then we see that the fixed point thus obtained is scale invariant but with a non-trivial Virial current:
\begin{align}
\tilde{T}^{\mu}_{\mu} &= \partial^\mu J_\mu \cr
J_\mu &=  J_\rho^{[IJ]} \delta_{I}^\rho \delta_{J\mu}  
\end{align}
Generically, the Virial current $J_\mu$ is not written as a derivative of symmetric tensors in a local manner $J_\rho = \partial^\nu L_{\nu\rho}$, so the non-trivial fixed point if any is only scale invariant but not conformal invariant \cite{Wess}\cite{Coleman:1970je}\cite{Polchinski:1987dy}. It is crucial that the zero of the beta function is non-trivial so that $O(d)_{\mathrm{internal}}$ is broken here. Thus we cannot undone the topological twist at such renormalization group fixed points.

\section{Field theory examples}
As one of the simplest examples of topologically twisted deformations, let us consider the three-dimensional free massless boson with the $O(d)$ global symmetry.
\begin{align}
S = \int d^d x \partial_\mu \Phi^I \partial^\mu \Phi^I \ .
\end{align}
The idea of topological twist is to identify the global $O(d)_{\mathrm{internal}}$ index $I$ with the space index $\mu$ so that we regard $\phi^I \delta^\mu_I$ as a vector under the topologically twisted rotation. The twisted action becomes
\begin{align}
S = \int d^dx \partial_\mu \Phi^\nu \partial^\mu \Phi_\nu \ . 
\end{align}
Note that we now treat $\Phi^\mu = \delta^{\mu}_I \Phi^I$ as a vector under the new rotational symmetry.

The original energy-momentum tensor after renaming was
\begin{align}
T_{\mu\nu} = -\partial_\mu \Phi_\rho \partial_\nu \Phi^\rho + \frac{\delta_{\mu\nu}}{2} \partial_\rho \Phi^\sigma \partial^\rho \Phi_\sigma \ .  
\end{align}
On the other hand, the topologically twisted energy-momentum tensor is computed as 
\begin{align}
\tilde{T}_{\mu\nu} &= - \partial_\mu \Phi^\rho \partial_\rho \Phi_\nu - \partial_\rho \Phi_\mu \partial_\nu \Phi^\rho + \partial_\mu \Phi^\rho \partial_\nu \Phi_\rho  + \partial_\rho \Phi_\mu \partial^\rho \Phi_\nu  + (\Phi_\mu \partial_\nu \partial^\rho \Phi_\rho + \Phi_\nu \partial_\nu \partial^\rho \Phi_\rho) \cr
& + \frac{\delta_{\mu\nu}}{2} (\partial_\rho \Phi_\sigma \partial^\rho \Phi_\sigma - \partial_\rho \Phi^\sigma \partial_\sigma \Phi^\rho - 2 \Phi^\rho \partial_\rho \partial^\sigma \Phi_\sigma - (\partial^\rho \Phi_\rho)^2 ) \ .
\end{align}
Up to the improvement terms and the use of the equations of motion $\partial^\rho \partial_\rho \Phi^\mu = 0$, they differ by the amount given in \eqref{newEM}.

Now we introduce the topologically twisted deformations that are singlet under the  twisted rotation. In generic space dimensions $d$, the original theory has a  spin one operator with the scaling dimension $\Delta = d$ in the vector representation i.e. $(\Phi^I \Phi^I) \partial_\mu \Phi_J$ and a spin two operator in the symmetric traceless tensor representation with the scaling dimension $\Delta_d$ (as well as the anti-symmetric tensor operators with the same dimension) i.e. $\partial_\mu \Phi^I \partial_\nu \Phi^J$, so one may consider the topologically twisted deformation:
\begin{align}
\delta S = \int d^dx  g_1 (\Phi^J \Phi^J) \partial_\mu \Phi^I \delta_{I}^\mu + g_2\partial_\mu \Phi^I \partial_\nu \Phi^J \delta_{I}^{\mu} \delta_{J}^{\nu} + g_3 \partial_\mu \Phi^I \partial_\nu \Phi^J \delta_{J}^{\mu} \delta_{I}^{\nu}   \ . 
\end{align}
The role of the deformations from the tensor  operators is to change the form of the kinetic term for the spin one field $\Phi^\mu$, so they are exactly marginal (without introducing the other terms). It is well known that such a  twisted theory is scale invariant without conformal invariance \cite{ElShowk:2011gz} (except for a very particular parameter for the kinetic term). Note that once we add this term to the action, we are no longer able to undone the topological twist.

As we have discussed, the situation in $d=3$ dimensions is special. There is additional universal deformation coming from the $O(3)$ current operator itself:
\begin{align}
\delta S =  \int d^3x J_\mu^I \delta^\mu_I = \int d^3x \epsilon_{\mu\nu\rho} \Phi^\mu \partial^\nu \Phi^\rho \ .
\end{align}
This operator has the scaling dimension $d-1 = 2$, and it is relevant. Once we regard $\Phi^\mu$ as a spin one field, the topologically twisted deformation may be identified with the Chern-Simons term although we have no gauge symmetry here. Therefore, we expect that the resulting theory becomes gapped after the introduction of this deformation.

Let us briefly discuss the similar story in $O(3)$ symmetric critical  Heisenberg model in three dimensions. After the topological twist, there are a couple of candidates for the topologically twisted deformations. First of all, we have the universal deformations from the $O(3)$ conserved current. What about the other possibilities? As we have discussed, the reflection positivity at the critical Heisenberg fixed point demands that topologically twisted scalar deformations come only from spin one and spin two operators. We do not know the scaling dimensions of these operators exactly, but the current state of the art technology of the conformal bootstrap with the extremal functional method \cite{ElShowk:2012hu}\cite{El-Showk:2014dwa} gives us a reasonable estimate of the scaling dimensions.\footnote{As an open source code, one may use JuliBootS \cite{Paulos:2014vya} for this purpose. We have used the customized version of cboot \cite{cboot} based on SDPB \cite{Simmons-Duffin:2015qma}.}
 It turns out that the second lowest spin one operators in the adjoint representations of $O(3)_{\mathrm{internal}}$ has the scaling dimension $\Delta_{1,A}' \sim 4.1$ and the lowest spin two operators in the symmetric traceless representations of  $O(3)_{\mathrm{internal}}$ has the scaling dimension $\Delta_{2,T} = 3.2$, so the conformal bootstrap suggests that there is no other topologically twisted scalar deformations that are relevant in the $O(3)$ symmetric critical Heisenberg model.


\section{Holographic realization}
Within the quantum field theory analysis, it is often difficult to follow the renormalization group flow induced by the topologically twisted operators, and the fate of such flow is not trivial. 
Instead of studying the flow in each systems, we would like to study the generic features of the flow in a certain class of theories realized by holography. 
In this section we construct a holographic dual description of the renormalization group flow induced by the topologically twisted operators by using the ansatz motivated by the cosmological models studied in the literature some time ago\cite{Henneaux:1982vs}\cite{Galtsov:1991un}.\footnote{There are some renewed interest in this ansatz e.g. in gauge-flation \cite{Maleknejad:2011jw}\cite{Maleknejad:2011sq} or chromo-natural inflation \cite{Adshead:2012kp}. We should stress, however, that our solution presented below do not survive in the de-Sitter space-time without violating the reality conditions of the $O(3)$ gauge field.}

As a minimal setup, we need the holographic counterpart of the energy-momentum tensor, the $O(d)$ conserved current and a topologically twisted scalar operator.  For the corresponding  minimal setup in the holographic side, we consider an $O(d)$ gauge theory coupled with the $d+1$ dimensional (Einstein) gravity. We may further add vector (or tensor) fields charged under $O(d)$ in the bulk that play the role of the topologically twisted scalar operators.\footnote{The role of tensor fields to violate the Lorentz invariance in the context of holography has been studied in \cite{Kiritsis:2012ta}.} In $d=3$, we have the universal deformations by the $O(3)$ conserved current itself, so the latter may not be necessary.

We assume that before the topologically twisted deformation, we have the bulk solution of the $d+1$ dimensional AdS space with the Euclidean Poincar\'e metric:
\begin{align}
ds^2 = \frac{dz^2 + \delta_{\mu\nu} dx^\mu dx^\nu }{z^2}
\end{align} 
with no non-trivial $O(d)$ gauge field configuration. The configuration preserves the isometry of $SO(d+1,1)$ that is identified with the Euclidean conformal group of the dual Euclidean field theory.

After the topologically twisted deformations, we are going to solve the equations of motion with the ansasz for the bulk metric
\begin{align}
ds^2 = f(z) \frac{dz^2 + \delta_{\mu\nu} dx^\mu dx^\nu}{z^2}
\end{align}
and the bulk vector field
\begin{align}
A^a = g(z) dx^i \delta_{i}^a \ .
\end{align}
When $d=3$, we may identify the vector field with the $O(3)$ gauge field itself. The non-trivial profile of $f(z)$ breaks the isometry corresponding to the scale symmetry, which describes the renormalization group flow.

If the theory admits the non-trivial fixed point for the renormalization group flow  induced by the topologically twisted scalar deformations, there must exist a solution of the equations motion with the Euclidean AdS background\footnote{As noted in \cite{Nakayama:2009qu}\cite{Nakayama:2010wx}\cite{Nakayama:2010zz} even without conformal invariance, the scale invariance alone dictates that the metric must take the form of AdS.}
\begin{align}
ds^2 = \frac{dz^2 + \delta_{\mu\nu} dx^\mu dx^\nu}{z^2} \label{sol1}
\end{align}
and the $O(d)$ vector field condensation with the form
\begin{align}
A^a = c\frac{dx^i \delta_{i}^a}{z} \ , \label{sol2}
\end{align}
where $c$ is a non-zero constant.
Generically, we expect that such a solution requires fine-tuning of the bulk equations of motion because two of the three equations, (i.e. $(zz)$ and $(ij)$ components of the Einstein equation and the Yang-Mills equation) are independent and we have to solve  the two equations only with one adjustable variable $c$. The situation should be contrasted with the usual scalar flow, in which the extrema of the scalar potential automatically solves the Einstein equation with the AdS background and we need to solve only one equation with one variable.

Therefore, we suspect that the holographic renormalization group flows with the topologically twisted deformations will end up with the confinement rather than reaching another non-trivial renormalization group fixed point with scale invariance without conformal invariance.
However, a miracle happens even in the simplest model.

Consider the $O(3)$ Yang-Mills theory coupled with the Einstein gravity in four dimensions with the minimal action
\begin{align}
 S = \int d^4x \sqrt{g} \left( R +6 + \mathrm{Tr} F_{\mu\nu}F^{\mu\nu} \right) .
\end{align}
Here we do not introduce the other vector fields so we are studying the universal deformations available only in the three dimensional dual conformal field theory.

It turns out that despite the general suspicion above, this theory does support the AdS geometry with the $O(3)$ gauge field configuration given by \eqref{sol1} and \eqref{sol2} with $c=1$.
A short calculation tells that both the Yang-Mills equation and the Einstein equation are simultaneously solved with the ansatz \eqref{sol2}. The easiest way to convince ourselves of the statement is to use the Weyl invariance of the Yang-Mills equation in four-dimensions to map the problem to the flat space. Then use the $\phi^4$ theory ansatz \cite{Corrigan:1976vj}\cite{Actor:1979in} 
\begin{align}
& A_z^a = \mp \delta^a_i \frac{{\partial^i} \phi}{\phi} \ , \ \  A_i^a = \epsilon_{ian} \partial_n \phi \pm \delta_{ai}\frac{\partial_z \phi}{\phi} \cr
& \partial_\mu \partial^\mu \phi + \lambda \phi^3 = 0  \ ,
\end{align}
where $\lambda$ is an arbitrary integration constant, to confirm the solution $\phi = 1/z$ with the vanishing Yang-Mills energy-momentum tensor. We find that the field strength is not self-dual (nor anti-self dual) because $\lambda \neq 0$.
The flat space solution with vanishing energy-momentum tensor can be mapped to the AdS space with no back reaction to the AdS metric, solving the all the components of the Einstein equation. 

The resulting field configuration is invariant under the twisted rotation (i.e. the diagonal of $O(3)$ gauge transformation and the $O(3)$ rotation) as well as scale transformation. However, one may see that it is not invariant under the original isometry that would correspond to the special conformal transformation.
In this way, we have explicitly constructed a holographic dual description of scale invariant but non-conformal field theories induced by the topologically twisted deformations.

\section{Discussions}
Our  study of the topologically twisted deformations of Euclidean field theories is just the beginning, and there are many directions to go. It is important to understand the properties of the renormalization group flow induced by the topologically twisted scalar deformations in various interacting quantum field theories. Are there any analogue of $c$-theorem? How can we compute the beta functions and their non-trivial zeros systematically?\footnote{The so-called vector beta functions \cite{Nakayama:2013ssa} before the topological twist directly give the beta functions of the topologically twisted operators.}
What would be the consistency conditions on the local renormalization group flow? 

Toward a main application of the topological twist in supersymmetric field theories, it is important to address the classification of the supersymmetric deformations of the superconformal field theories under the topologically twisted deformations. Such a classification without the topological twist has been throughly discussed in a recent paper \cite{Cordova:2016xhm}, but we have seen that we have more deformations if we allow the topological twist.

Our holographic realization is ubiquitous and may be embedded in supergravity or string theory. The natural question, then, is what would be the dual field theory interpretations of such a universal background? More physically, it may be important to revisit Euclidean statistical models studied in the literature to check if we might not overlook such possibilities and the fixed points we had obtained might or might not be affected by topologically twisted deformations.

Our solution strongly suggests that there are infinitely many scale invariant but not conformally invariant Euclidean field theories that can be constructed as an end point of the renormalization group flow induced by topologically twisted deformations. The general properties of such theories are poorly understood compared with the conformal field theories. Once we found the black swan, the  new world is welcoming us.

\section*{Acknowledgements}
The author would like to thank I.~Obata for his lecture on his paper \cite{Obata:2016tmo} at Rikkyo university, which was the main inspiration of the paper. He acknowledges T. Ohtsuki for his dedication to the conformal bootstrap code (as well as his endurance to the author's complaint), a part of which was used in this paper.

\end{document}